\documentclass[aps,pra,showpacs,twocolumn,superscriptaddress,longbibliography]{revtex4-1}
\usepackage{graphicx}
\usepackage[usenames]{color}
\usepackage{amssymb,amsmath}
\usepackage{siunitx}%To use degree symbol
\usepackage{xcolor}
\usepackage{setspace} 
\usepackage{float} 
\usepackage{tabularx}
\newcommand{\eq}[1]{Eq.~(\ref{#1})}
\newcommand{\fig}[1]{Fig.~\ref{#1}}
\newcommand{\be}[1]{\begin{equation}\label{#1}}
\newcommand{\ee}{\end{equation}}

\usepackage[LGRgreek]{mathastext}
\usepackage{lipsum}

\begin{document}

\title{Enhancing   frustrated double ionisation with no electronic correlation in triatomic molecules using counter-rotating two-color circular laser fields }

\author{G. P. Katsoulis}
\affiliation{Department of Physics and Astronomy, University College London, Gower Street, London WC1E 6BT, United Kingdom}
\author{R. Sarkar}
\affiliation{Department of Physics and Astronomy, University College London, Gower Street, London WC1E 6BT, United Kingdom}
\author{A. Emmanouilidou}
\affiliation{Department of Physics and Astronomy, University College London, Gower Street, London WC1E 6BT, United Kingdom}

\begin{abstract}

We demonstrate significant enhancement of frustrated double ionization (FDI) in the two-electron triatomic molecule D$_{3}^{+}$ when driven by counter-rotating two-color circular (CRTC) laser fields. We employ a three-dimensional semiclassical model that fully accounts 
for electron and nuclear motion in strong fields. For different pairs of wavelengths, we compute the probabilities of the FDI pathways  as a function of the ratio of the two field-strengths. We identify a pathway of frustrated double ionization  that is not present  in  strongly-driven molecules   with linear fields. In this pathway the first ionization step is ``frustrated" and electronic correlation is essentially absent. This pathway   is responsible for 
enhancing frustrated double ionization with CRTC  fields. We also employ a simple model that predicts   many of the main features   of the probabilities  of the FDI pathways   as a function of the  ratio of the two field-strengths.

 \end{abstract}
\pacs{33.80.Rv, 34.80.Gs, 42.50.Hz}
\date{\today}

\maketitle
\section{Introduction}
Formation of highly excited Rydberg states, during the interaction of  atoms and molecules with laser fields, is a fundamental problem with a wide range of  applications. Rydberg states underlie, for instance, acceleration of neutral particles \cite{Eichmann},  spectral features of photoelectrons \cite{Veltheim}, and formation of molecules via long-range interactions \cite{Bendkowsky}.  Recently, the formation of Rydberg states in weakly-driven  H$_{2}$  was accounted for by electron-nuclear correlated multi-photon resonant excitation   \cite{Zhang}. For  H$_2$ driven by intense infrared laser fields (strongly-driven), this latter process was shown to merge  with frustrated double ionization (FDI)  \cite{Zhang}. Frustrated double ionization accounts for the formation of Rydberg fragments in strongly-driven two-electron molecules. 
In frustrated ionization an electron first tunnel ionizes in the driving laser field. Then, due to the electric field, this electron is recaptured by the parent ion in a Rydberg state \cite{Nubbemeyer}. In frustrated double ionization an electron escapes while another one occupies a Rydberg state at the end of the laser pulse. 

For linear laser fields, frustrated double ionization  is a major process  during the breakup of  strongly-driven molecules, accounting for roughly 10\% of all ionization events. Hence, frustrated double ionization has been  the focus of intense experimental studies  in the context of H$_{2}$ \cite{Manschwetus}, D$_{2}$ \cite{Zhang2} and of the two-electron triatomic molecules D$_{3}^{+}$ and H$_{3}^{+}$ \cite{McKenna1, MScKenna2, Sayler}.   For strongly-driven two-electron diatomic and triatomic molecules, frustrated double ionization proceeds via two pathways \cite{Emmanouilidou,Emmanouilidou1,Emmanouilidou2}. One electron tunnel ionizes early on (first step), while  the remaining bound electron  does so later in time (second step). If  the second (first) ionization step is ``frustrated", we label the FDI pathway as FSIS (FFIS), i.e. ``frustrated" second (first) ionization step, previously referred to as pathway A (B)  \cite{Emmanouilidou}. Electron-electron correlation, underlying  pathway FFIS  \cite{Emmanouilidou, Emmanouilidou3}, can be controlled with orthogonally polarised two-color linear (OTC) laser fields  \cite{Emmanouilidou2}.

Here, we show that counter-rotating  two-color circular (CRTC)  laser fields are a powerful tool for controlling frustrated double ionization in strongly-driven molecules. CRTC fields have attracted a lot of  interest due to their applicability to the  production, via high harmonic generation,  of circular pulses with extreme-ultraviolet to  soft-x-ray wavelengths \cite{Eichmann2,Long, Milosevic, Bandrauk1, Bandrauk2}. This capability of CRTC fields has been demonstrated in groundbreaking experiments \cite{Fleischer, Kfir, Fan}. The latter open the way  to investigate chirality-sensitive light-matter interactions \cite{Bowering, Travnikova} and probe  properties of magnetic structures \cite{Radu,Flores}. Moreover, the relative intensity of the two colors in CRTC  fields has  been used to control nonsequential double ionization in driven atoms \cite{Chaloupka,Mancuso,Eckart} and molecules \cite{Wu_2017}.

We demonstrate that CRTC fields significantly enhance  frustrated double ionization  in D$_{3}^{+}$, compared to OTC fields \cite{Emmanouilidou2, Larimian}. Pathway FFIS accounts for the increase in the formation of Rydberg fragments.  
We find that electron-electron correlation does not necessarily underly  pathway FFIS. This is unlike  our findings with linear fields. If anything,  a significant enhancement of pathway FFIS coincides, roughly,  with  an absence of electronic correlation. 
 Hence,  pathway FFIS is a more general route to frustrated double ionization  than previously recognized \cite{Emmanouilidou,Emmanouilidou2}.
 We find that pathway FSIS and FFIS, the latter pathway  with or without electronic correlation,   prevail at different ratios of the field-strengths  of CRTC. Thus, appropriate tuning of the field strengths results in controlling the prevalent route to frustrated double ionization.  Importantly,  we employ a simple model that successfully accounts for many of the main features of frustrated double ionization and its pathways,  where the latter are obtained with a full-scale computation. 

We  focus on frustrated double ionization in D$_{3}^{+}$ driven by CRTC fields with wavelengths $\lambda_{1}=800$ nm and $\lambda_{2}=400$ nm. We achieve maximum enhancement  of frustrated double ionization when the ratio of the field strengths is  $\epsilon_{1}=\mathcal{E}_{2}/\mathcal{E}_{1}=4$. Frustrated double ionization accounts roughly for 20\% of all ionization events. We develop a  simple model to  explain    main features of the full-scale-computed probabilities of the  frustrated double ionization pathways as a function of $\epsilon_{1}$. Further below, we show that  this model predicts main features of frustrated double ionization  for a  range of pairs of wavelengths. We  set $\mathcal{E}_{1}+\mathcal{E}_{2}=0.08$ a.u., intensity of  2.25$\times 10^{14}$ W/cm$^2$,  keeping roughly constant the ionisation probability. In what follows, we employ atomic units unless otherwise stated. 
 
 \section{Method}
 We employ a three-dimensional (3D) semiclassical model for our full-scale computations \cite{Emmanouilidou1,Emmanouilidou3}. We choose an initial state of D$_{3}^{+}$ that is accessed experimentally via the reaction D$_{2}$ + D$_2^+$$\rightarrow$D$_{3}^{+}$ + D \cite{McKenna1,MScKenna2}. This state consists of a superposition of vibrational states $\nu =$ 1–12 with equilateral triangular-shape  \cite{MScKenna2, Anicich,Talbi}. Using the energy of each vibrational state from ref. \cite{Anicich}  and the potential energy curves as a function of inter-nuclear distance in ref. \cite{Talbi}, we obtain the inter-nuclear distance for each vibrational state, which varies from 2.04 a.u. ($\nu =$ 1) to 2.92 a.u. ($\nu =$ 12). We find the first and second ionization potentials for the 12 vibrational states using the quantum chemistry software MOLPRO \cite{molpro}. For each vibrational state, we initialise the three nuclei at rest, since  an initial predissociation does not significantly alter the ionization dynamics \cite{Emmanouilidou3}. 
Moreover, the strength of the combined  field  is  within the below-the-barrier ionization regime. Hence, one electron (electron 1) tunnel ionizes at time t$_{0}$ through the field-lowered Coulomb potential.  It does so with a rate given by  a quantum mechanical formula  \cite{Murray}. The exit point is taken along the direction of  the field  \cite{Emmanouilidou3}. The electron momentum  parallel  to the combined field is equal to zero. The transverse momentum is given by a Gaussian distribution   which represents the Gaussian-shaped filter with an intensity-dependent width arising from standard tunnelling theory \cite{tunn1,tunn2,tunn3}. 
The initially bound electron (electron 2) is described by a microcanonical distribution \cite{Chen}.

\begin{figure}[ht]
\includegraphics[scale=0.5]{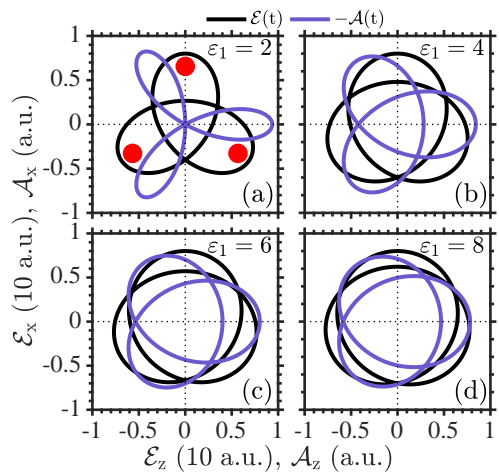}
\centering
\caption{Components of the electric field $\mathcal{E}$ and the vector potential $\mathcal{A}$ of the CRTC fields  for $\epsilon_{2}=2$, with field envelope set equal to 1. The red dots denote the nuclei in D$_{3}^{+}$.}
\label{fig0}
\end{figure}

 We use CRTC  fields of the form

 \begin{align}
  \vec{\mathcal{E}}(t) =&  \exp\left[ -2\ln 2 \left(\frac{t}{\tau}\right)^2 \right] \times \\ \nonumber
  &\left [ \mathcal{E}_1 (  \mathrm{\hat{x}}\cos \omega_1 \mathrm{t}  + \mathrm{\hat{z}}\sin \omega_1 \mathrm{t} ) +\mathcal{E}_2 (  \mathrm{\hat{x}}\cos \omega_2 \mathrm{t}  -\mathrm{\hat{z}}\sin \omega_2 \mathrm{t} ) \right],
  \label{eqnew}
 \end{align}
where $\tau=40$ fs is the full width at half maximum of the pulse duration in intensity. For the ratios $\epsilon_1=\mathcal{E}_{2}/\mathcal{E}_{1}$ and $\epsilon_2=\lambda_{1}/\lambda_{2}=\omega_{2}/\omega_{1}$ considered here, the combined laser field has $\epsilon_{2}+1$ lobes, see \fig{fig0}.
 Once the tunnel-ionisation time $t_{0}$  
 is selected randomly in the  time interval $[-2\tau,2\tau]$, we specify   
 the initial conditions. Then, employing the Hamiltonian of the strongly driven five-body system, we propagate classically the position and momentum of the electrons and nuclei. All Coulomb forces and the interaction of each electron and  nucleus with the CRTC fields are fully accounted for with no approximation. We also fully account for the Coulomb singularities \cite{Emmanouilidou3}. The motion of the electrons and the nuclei  are treated on an equal footing,   accounting for the interwind electron-nuclear  dynamics \cite{Emmanouilidou4,Zhang}. During propagation, we allow each electron to tunnel with a quantum-mechanical probability given by the Wentzel-Kramers-Brillouin approximation \cite{Emmanouilidou, Emmanouilidou3}.  We thus accurately account for enhanced ionization   \cite{Niikura,EI1,EI2,EI3,EI4}. In enhanced ionization, at a critical distance of the nuclei, a double-potential well is formed such that it is easier for an electron bound to the higher potential well to tunnel to the lower potential well  and then  ionize.  We note that the approximations considered in our model in the initial state and during the propagation are justified by the very good  agreement of our previous results for  H$_{2}$ \cite{Emmanouilidou} and D$_{3}^{+}$ \cite{Emmanouilidou1} with experimental results \cite{Manschwetus, MScKenna2}.

\section{Results}
 In frustrated double ionization of D$_{3}^{+}$ the final fragments are a neutral excited fragment D$^{*}$, two D$^{+}$ ions, and one escaping electron. In the neutral excited fragment D$^{*}$ the electron transitions to a Rydberg state with quantum number n$ >$1. Here, we find that the Rydberg state with $n\approx 10$ is the  most probable to form during frustrated double ionization with CRTC fields.  In pathway FSIS, electron 1 tunnel ionizes and escapes early on. Electron 2 gains energy from an enhanced-ionization-like process and tunnel ionizes. However, it does not have enough drift energy to escape when CRTC is turned off, and occupies a Rydberg state, D$^{*}$.    In pathway FFIS, electron 1 tunnel ionizes and quivers in the laser field. Electron 2  tunnel ionizes after a few periods of the laser field. Electron 2 gains energy from an enhanced-ionization-like process. Depending on $\epsilon_{1}$ and $\epsilon_{2}$, electron 2 can, in addition, gain energy from the returning electron 1 via electron-electron correlation.   When the laser field is turned off, electron 1 does not have enough energy to escape and remains bound in a Rydberg state. In studies with linear laser fields, we  found that electronic correlation underlies  pathway FFIS \cite{Emmanouilidou,Emmanouilidou3}. For CRTC fields,  we show that electronic correlation  underlies pathway FFIS only for certain $\epsilon_1$ values.

  \begin{figure*}[ht]
\begin{centering}
\includegraphics[scale=0.45]{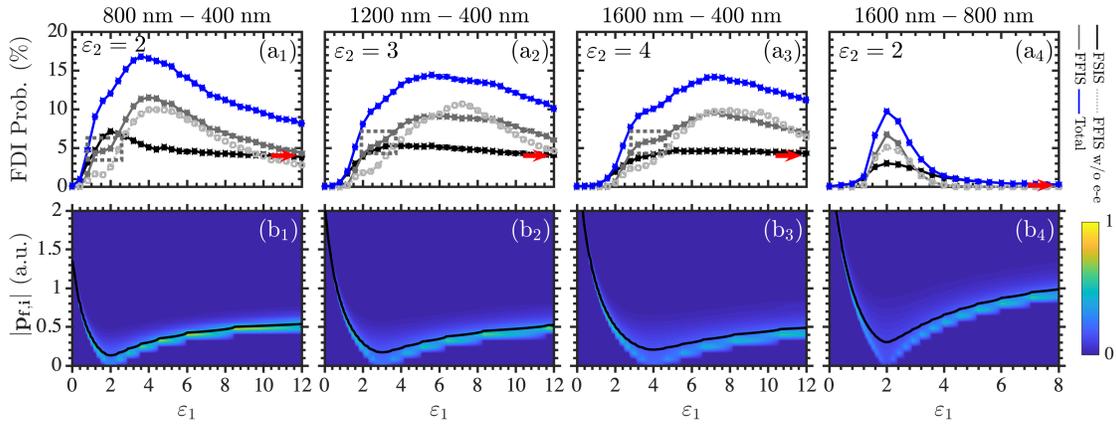}
\caption{For different sets of $\lambda$s with $\mathcal{E}_{1}+\mathcal{E}_{2}=0.08$ a.u., for D$_{3}^{+}$, we plot as a function of $\epsilon_{1}$ (top row)  the  FDI probabilities, computed using the full-scale 3D model;  (bottom row) the probability distribution of p$_{f, i}$ and its mean value (black line) computed using the simple model. The red arrows indicate the FDI probability when $\mathcal{E}_{1}=0$ a.u. The light grey curves in the top row labeled FFIS w/o e-e correspond to the FFIS probabilities when electron-electron correlation is turned off. }
\label{fig1}
\end{centering}
\end{figure*}

We compute the FDI probability using 
\begin{equation}
P^{FDI}(\epsilon_1,\epsilon_2)=\frac{\sum_{\nu,i}P_{\nu}\Gamma(\nu,i,\epsilon_1,\epsilon_2)P^{FDI}(\nu,i,\epsilon_1,\epsilon_2)}{\sum_{\nu,i}P_{\nu} \Gamma(\nu,i,\epsilon_1,\epsilon_2)},
\label{eqprob}
\end{equation}
 where $i$ denotes the different orientations of the molecule. $\Gamma(\nu,i,\epsilon_1,\epsilon_2)$ is given by
 \begin{equation}
 \Gamma(\nu,i,\epsilon_1,\epsilon_2)=\int_{t_{in}}^{t_{f}}\Gamma(t_{0},\nu,i,\epsilon_1,\epsilon_2)dt_{0}, 
 \end{equation}
 where the integration is over the duration of CRTC and  $\Gamma(t_{0}, \nu,i,\epsilon_1,\epsilon_2)$ is the ionization rate. $\Gamma(\nu,i,\epsilon_1,\epsilon_2)$ remains  roughly the same, for constant  $\mathcal{E}_1+\mathcal{E}_2$.
  The percentage of the vibrational state $\nu$ in the initial state of D$_{3}^{+}$ \cite{Anicich} is denoted by  P$_{\nu}$.  The probability $P^{FDI}(\nu,i,\epsilon_1,\epsilon_2)$ is the number of FDI events out of all initiated classical trajectories.
  The computations involved are challenging. We approximate 
 \eq{eqprob} using the $\nu=8$ state, which we find to contribute the most in the sum in \eq{eqprob}.  For CRTC fields with $\lambda_{1}=$800 nm and $\lambda_{2}=$400 nm, we obtain very similar results for the $\nu=6,7,8$ states. Since the $\nu=6,7$ states contribute less to the sum in \eq{eqprob} than the $\nu=8$ state but more than the other states, we approximate 
 \eq{eqprob} using the $\nu=8$ state. We expect this to be the case for all other pairs of wavelengths considered in the current work. Moreover, for CRTC fields with $\lambda_{1}=$800 nm and $\lambda_{2}=$400 nm,
  we consider two planar alignments, with one side of the molecular triangle being either parallel or perpendicular to the x-axis, the latter is shown in \fig{fig0}(a). 
 We find that  the change of P$^{FDI}$ with  $\epsilon_1$ is roughly the same for both orientations and expect this to be the case  for all other orientations. Thus, we choose the perpendicular orientation to compute our  results for all other pairs of wavelengths considered in this work.
  We note that changing the phase between the components of the electric field in Eq. (1), which currently is set to be equal to zero, only changes the planar alignment of the electric field with respect to the molecule. Hence, we expect that  the probability for frustrated double ionization is not affected by this phase.

 For CRTC fields with $\lambda_{1}=$800 nm and $\lambda_{2}=$400 nm, the dependence on $\epsilon_{1}$  of the total FDI probability   and of  the FSIS and FFIS probabilities have several interesting features, see \fig{fig1}(a1). The FDI probability reaches 17\%   at $\epsilon_{1}\approx4$. This is twice the  FDI probability we  computed previously  for both a single linear pulse of 800 nm \cite{Emmanouilidou1} and an OTC  pulse with 800 nm and 400 nm  \cite{Emmanouilidou2}. Thus, CRTC fields  significantly enhance  frustrated double ionization in D$_{3}^{+}$. In studies with linear fields, pathways FSIS and FFIS contribute to frustrated double ionization roughly the same  \cite{Emmanouilidou1,Emmanouilidou2}. In contrast, for CRTC fields, the FFIS probability is roughly twice the FSIS probability for $\epsilon_{1}\approx4$, see \fig{fig1}(a1).   
 
 Another striking feature of the change of the FDI probability with  $\epsilon_{1}$,  is  the ``plateau"  the FFIS probability exhibits around $\epsilon_{1}=2$.  For smaller  $\epsilon_{1}$ and larger values up to  $\epsilon_{max}^{FFIS}$, the FFIS probability increases sharply.
  The value  $\epsilon_{max}^{FSIS}$ ($\epsilon_{max}^{FFIS}$)   corresponds to the peak of  the  FSIS (FFIS) probability. A ``plateau"  suggests a different mechanism underlying pathway FFIS  at  $\epsilon_{1}=2$ compared to other $\epsilon_1$  values. Indeed, we find that electronic correlation plays a major role in pathway FFIS mostly around $\epsilon_{1}=2$. To show this, we also compute the FDI probabilities  with   electron-electron correlation turned off in our 3D semiclassical model, see light grey curve in \fig{fig1}(a1-a3) labeled FFIS w/o e-e. Comparing the FFIS probabilities with and without electronic correlation, we find that the FFIS  probability reduces  by more than 50\%  around $\epsilon_{1}=2$, see \fig{fig1}(a1). However, the effect of electron-electron correlation is small on the FFIS probability for values of $\epsilon_{1}$ larger than 2. The FSIS probability 
   remains roughly the same (not shown), as is the case for linear fields \cite{Emmanouilidou1,Emmanouilidou2}.    Moreover,  compared to the FSIS probability,  FFIS peaks at a higher $\epsilon_1$  and reduces at a much faster rate for large $\epsilon_{1}$, see \fig{fig1}(a1). 
       
To understand these features of the change of the FDI probabilities with $\epsilon_{1}$, we employ a simple model.  This model entails an  estimate   of the final electron momentum   in the presence of  the CRTC fields, defined in Eq. 1. This momentum   largely  determines whether an electron finally escapes or occupies a Rydberg state. Neglecting electronic correlation, conservation of energy gives
\begin{equation}
\frac{({\bf p}_{i}(t_{ion})-{\bf \mathcal{A}}(t_{ion}))^2}{2}-V(r_{i,1},r_{i,2},r_{i,3})=\frac{{\bf p}_{i}(t\rightarrow\infty)^2}{2}=\frac{{\bf p}_{f,i}^2}{2},
\label{eq2}
\end{equation}
where t$_{ion}$ is the ionization time of an electron $i=1,2$, defined as the time when the compensated energy  becomes positive and remains positive thereafter \cite{comp}; ${\bf p}_{i}=p_{x,i}\hat{x}+p_{y,i}\hat{y}+p_{z,i}\hat{z}$, $r_{i,j}$ is the distance of electron i from nucleus j, with $j=1,2,3$, and V is the Coulomb interaction of  electron i with the nuclei. We further simplify and set  $V\approx0$. We also set  the electron momentum at the time of ionization p$_{i}(t_{ion})\approx0$. Then, 
the final momentum ${\bf{p}}_{f,i}$ is given by -$\mathcal{A}(t_{ion})$,  see purple lines in \fig{fig0}.  We note that in this simple model the momentum of each electron is the same, that is, ${\bf{p}}_{f,1}={\bf{p}}_{f,2}= -\mathcal{A}(t_{ion})$.  Using these assumptions and setting the field envelope equal to 1, we solve the classical equations of motion to obtain 
\begin{align}
\begin{split}
{\bf p}_{i} (t)=&\mathcal{A}(t)-\mathcal{A}(t_{ion})\\
{\bf r}_{ i}(t)=& {\bf r}_{ i} (t_{ ion })+\int_{t_ {ion}}^{t} \mathcal{A}(t')dt'-\int_{t_{ion}}^{t} \mathcal{A}(t _{ion})dt'.
\end{split}
\end{align}
We  find that   an electron returns to its initial position at the time of ionization, i.e.  ${\bf r}_{ i}(t)={\bf r}_{ i} (t_{ ion })$, at times  $\omega_{1}t=2(n+1)\pi$ if $\omega_{1}t_{ion}=2n\pi$ and $\epsilon_{1}=\epsilon_{2}$, with n an integer.  For these conditions, ${\bf p}_{i}(t)$ and ${\bf p}_ {f,i} $  are  zero. 
  We note that the prediction of our simple model that when $\epsilon_{1}=\epsilon_{2}$ the energy of either electron is very small and either electron returns to the molecular core are consistent with experiments  for CRTC fields with $\lambda_{1}=$800 nm and $\lambda_{2}=$400 nm \cite{Mancuso2016}.
  Namely, in ref. \cite{Mancuso2016}, the three-dimensional photoelectron distributions for atoms driven by CRTC fields are measured. It is found that when $\mathcal{E}_{2}=2\mathcal{E}_{1}$ the electrons are driven to a minimum in the momentum spectrum and also that the number of electrons in close-proximity to the parent ion  reaches a maximum. 
  
 Next, we obtain, for each $\epsilon_{1}$, the distribution of the magnitude of the final electron momentum ${\bf p}_{f,i}$. Namely,  we  compute $|\mathcal{A}(t_{ ion})|$ in the time interval t$_{ion}\in [0, T)$, with T the period of the CRTC fields. An electron tunnel ionizes with  different rates at different times.
To account for this, we   take the  tunnel ionization and ionization times to be the same. This assumption is more accurate for electron 1. For simplicity, we assume that electron i ionizes with 
  the atomic quantum tunnelling rate $\Gamma_{ADK}$ given by the Ammosov-Delone-Krainov (ADK) formula \cite{tunn1}. In $\Gamma_{ADK}$, we set the ionization energy I$_{p}$ equal to the first ionization potential of D$_{3}^{+}$. We take  the effective charge Z$_{eff}$ equal to the asymptotic one an electron ``sees" when moving away from  D$_{3}^{+}$. Our results are similar for other values of I$_{p}$ and Z$_{eff}$. Weighting  each p$_{f,i}$ by  $\Gamma_{ADK}(t_{ion})/\int_{0}^{T}\Gamma_{ADK}(t_{ion})dt_{ion}$, we obtain the distribution of p$_{f,i}$ as a function of $\epsilon_{1}$, shown in \fig{fig1}(b1).

Below we show that  the change of the distribution of p$_{f,i}$ with $\epsilon_{1}$ accounts for the main features of the frustrated double ionization  probabilities as a function of $\epsilon_1$. 
The distribution of p$_{f,i}$,   for  $0<\epsilon_{1}<\epsilon_{2}$, is narrow and decreases sharply,   starting from large values and reaching  zero at $\epsilon_{1}=\epsilon_{2}$,  see \fig{fig1}(b1). For $\epsilon_{1}>\epsilon_2$, the distribution of p$_{f,i}$ is wide and increases   at a slower rate than its decrease rate for $\epsilon_{1}<\epsilon_{2}$, reaching at $\epsilon_1\rightarrow \infty$  smaller values than at $\epsilon_{1}=0$. For large p$_{f,i}$ values,  either electron has a high chance of finally escaping at the end of the laser field, i.e. a small chance for frustrated double ionization. Indeed, the FDI probabilities are roughly zero at $\epsilon_{1}=0$ (\fig{fig1}(a1)). 
   Moreover,    decreasing (increasing) p$_{f,i}$ values result in either electron having an increasing (decreasing)  chance of remaining bound at the end of the  laser field. This in turn implies   an increasing (decreasing)  chance of frustrated double ionization. A comparison of  \fig{fig1}(a1) with \fig{fig1}(b1) shows that, indeed, the total FDI probability  increases (decreases)   
  when p$_{f,i}$  decreases (increases). Also,  both the FDI probability and the distribution of p$_{f,i}$ are changing with a similar rate with $\epsilon_{1}$, albeit the sign of change.  When  the values of the distribution of the magnitude of the final electron momentum $p_{f,i}$ at $\epsilon_{1}=0$ are much larger than the values of $p_{f,i}$ at  $\epsilon_{1}\rightarrow \infty$, see  \fig{fig1}(b1)-(b3),  we find that the FDI probability is larger  at $\epsilon_{1}\rightarrow \infty$ than at $\epsilon_{1}=0$, see \fig{fig1}(a1)-(a3). However, if the values of $p_{f,i}$ at $\epsilon_{1}\rightarrow \infty$ become large, see   \fig{fig1}(b4), then the FDI probability will be zero both at  $\epsilon_{1}\rightarrow \infty$ and at $\epsilon_{1}=0$, see  \fig{fig1}(a4).

 Next, we explain why the FFIS probability  reduces at a much faster rate compared to FSIS, for  $\epsilon_{1}>\epsilon_{max}^{FFIS}$, see \fig{fig1}(a1). The Coulomb interaction of each electron with the nuclei is ignored in our simple model, it is however present in  the full-scale 3D semiclassical  model. Analysis of the results,  we obtain with the full-scale model, shows that electron 1  tunnel-ionises  further away from the nuclei compared to electron 2.  For large $\epsilon_{1}$, the slow increase of p$_{f,i}$  results mostly in the less bound electron having an increased chance to escape the Coulomb potential and thus ionize. Hence, electron 1 in pathway FFIS has a higher chance to escape  compared to   electron 2 in FSIS, reducing the FFIS probability at a faster rate.

 Returning to the  ``plateau" in the FFIS probability, we further discuss the underlying mechanism. As explained, at  $\epsilon_{1}=\epsilon_{2}$,   p$_{f,i}$ reaches zero. Small p$_{f,i}$ values   mostly inhibit the final escape of the more tightly bound electron 2. Thus,  in order for frustrated double ionization to proceed via pathway FFIS     an extra energy transfer from electron 1 to electron 2 is required. This is provided by the strong electron-electron  correlation resulting from the return of electron 1 to the nuclei at $\epsilon_{1}=\epsilon_{2}$, as discussed earlier. This is in accord with  the ``plateau" in the FFIS probability being around $\epsilon_{1}=\epsilon_{2}$ for CRTC fields with $\lambda_{1}=800$ nm and $\lambda_{2}=400$ nm. Hence, for CRTC fields, our findings  suggest that electron-electron correlation underlies pathway FFIS mostly when p$_{f,i}$ has small values. This is corroborated by our previous finding of electron-electron correlation underlying pathway FFIS for linear fields \cite{Emmanouilidou1,Emmanouilidou2}. Indeed, for a linear field with $\mathcal{E}=0.08$ a.u. and $\lambda=800$ nm, the simple model we employ yields small p$_{f,i}$ values. 
  \begin{figure*}[ht]
\begin{centering}
\includegraphics[scale=0.8]{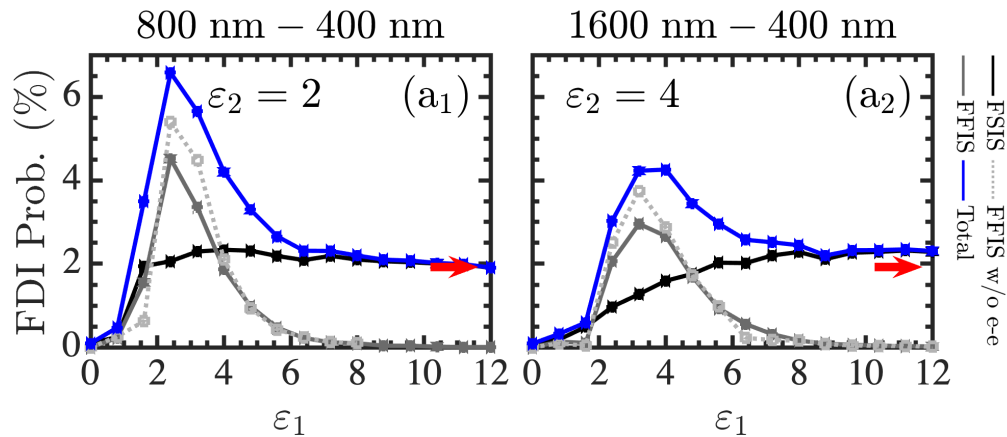}
\caption{For two sets of $\lambda$s with $\mathcal{E}_{1}+\mathcal{E}_{2}=0.064$ a.u., for H$_{2}$, we plot as a function of $\epsilon_{1}$ (top row)  the  FDI probabilities, computed using the full-scale 3D model. }
\label{fig3}
\end{centering}
\end{figure*}

The simple  model described above  predicts the change of  the FDI probabilities with $\epsilon_{1}$ for a range of  pairs of  wavelengths  of CRTC fields.
Indeed,   first, using this simple model, we compute the distribution of p$_{f,i}$ with  $\epsilon_{1}$ for  $\lambda_{1}=1200$ nm and $\lambda_2=400$ nm, i.e. $
\epsilon_{2}=3$ (\fig{fig1}(b2)), for $\lambda_1=1600$ and $\lambda_{2}=400$ nm,  i.e. $\epsilon_{2}=4$ (\fig{fig1}(b3)), and for $\lambda_{1}=1600$ nm and $\lambda_{2}=800$ nm, i.e $\epsilon_{2}=2$  (\fig{fig1}(b4)). Next, for the same parameters, using our full-scale 3D semiclassical  model,  we  compute the probabilities of the total frustrated double ionization and its pathways in \fig{fig1}(a2)-(a4).  As expected,  a comparison of \fig{fig1}(a2)-(a4) with  \fig{fig1}(b2)-(b4), respectively, reveals that  electron-electron correlation underlies pathway FFIS mainly at $\epsilon_{1}\approx\epsilon_{2}$, see  ``plateau"  enclosed by the dotted square in \fig{fig1}(a1)-(a3). This is the case for all pairs of $\lambda$s with small $p_{f,i}$ around $\epsilon_{1}=\epsilon_{2}$, \fig{fig1}(b1)-(b3).  In contrast for  $\lambda_{1}=1600$ nm and $\lambda_{2}=800$ nm, $p_{f,i}$ is not as small around $\epsilon_{1}=\epsilon_{2}$ (\fig{fig1}(b4)), resulting in electronic correlation having a  small effect in FDI probabilities for all $\epsilon_{1}$s (\fig{fig1}(a4)).
 Also, as discussed above, for large $\epsilon_{1}$, 
  the FFIS probability decreases in accord with the  increase of  p$_{f,i}$ with $\epsilon_{1}$. Indeed,  the rate of decrease of the FFIS probability  is    higher  for  $\lambda_{1}=1600$ nm and $\lambda_{2}=800$ nm compared to $\epsilon_{2}=3,4$ with $\lambda_{2}=400$ nm, since p$_{f,i}$ increases at a faster rate in the former case. 
  
  We note that while we find maximum enhancement of frustrated double ionization in D$_{3}^{+}$ driven by CRTC fields with wavelengths 800 nm and 400 nm, the FDI probability is also enhanced for wavelengths 1200 nm and 400 nm (see \fig{fig1}(a2)) as well as for 1600 nm and 400 nm (see \fig{fig1}(a3)).  Hence, further studies are needed with molecules of different symmetries to find whether maximum enhancement of frustrated double ionization is achieved with CRTC fields with symmetry that is closest to the symmetry of the initial molecular state.

 Which pair of $\lambda$s  is best suited to  infer experimentally   electronic correlation in frustrated double ionization in D$_{3}^{+}$ driven by CRTC fields? One expects  to be  the pair of $\lambda$s  resulting in p$_{f,i}$ being roughly zero over a wider  region of $\epsilon_{1}$. This condition is best satisfied for    $\lambda_1=1600$
and $\lambda_{2}=400$ nm, see \fig{fig1}(b3), giving rise  to a wider, and thus more visible,  ``plateau"  in the FFIS probability, see \fig{fig1}(a3).

We now show that pathway FFIS with no electron-electron correlation  is also present in frustrated double ionization of H$_{2}$ when driven by CRTC fields, see \fig{fig3}(a1)-(a2). We choose $\mathcal{E}_{1}+\mathcal{E}_{2}=0.064$ a.u. so that the sum of the field strengths of 0.064 a.u. for H$_{2}$ and 0.08 a.u. for D$_{3}^{+}$
has the same percentage difference from the field strength that corresponds to over-the-barrier ionization. We find that the FDI probability is around 6\% for $\epsilon_{2}=2$, which is roughly the same with the FDI probability for a linear laser field of wavelength 800 nm along the axis of the molecule. That is, unlike D$_{3}^{+}$, enhancement of frustrated double ionization is not achieved for H$_{2}$ with CRTC fields.  We also find that  the FFIS probability for H$_{2}$ reduces at a much faster rate compared to FFIS for D$_{3}^{+}$. This is consistent with electron 1 experiencing a smaller Coulomb attraction from the nuclei  in H$_{2}^{+}$ (two nuclei) versus D$_{3}^{+}$ (three nuclei), compare \fig{fig1}(a1) with \fig{fig3}(a1). Moreover, unlike D$_{3}^{+}$, pathway FFIS  without electronic correlation  prevails for most $\epsilon_{1}$s for $\epsilon_{2}=2$ and even more so for $\epsilon_{2}=4$, see \fig{fig3}(a1)-(a2). In contract for D$_{3}^{+}$,   pathways FFIS with and without electronic correlation prevail at different  values of $\epsilon_{1}$ and are thus  well separated. This difference between D$_{3}^{+}$ and H$_{2}$ is consistent with the probability of pathway FSIS  peaking around $\epsilon_{1}=\epsilon_{2}$, and the one for FFIS peaking at larger  $\epsilon_{1}$ values for D$_{3}^{+}$, while it is the other way around  for H$_{2}$. We conjecture that this latter difference is related with our finding (not shown) that double ionization 
prevails when D$_{3}^{+}$ is driven by CRTC fields. In contrast,  for the field parameters considered in the current work, we find that single ionization prevails when H$_{2}$ is driven by CRTC fields. More studies are needed to verify whether this is indeed the case. In general, since, according to our simple model,  the smallest values of $p_{f,i}$ are around $\epsilon_{1}=\epsilon_{2}$, one expects that the FDI probability will be maximum around $\epsilon_{1}=\epsilon_{2}$.

\section{Conclusion}

We have shown that strong driving of two-electron triatomic molecules with counter-rotating two-color  circular laser fields significantly enhances frustrated double ionization compared to linear fields. For each pair of wavelengths, by suitably tuning the ratio of the two field-strengths, we  achieve significant enhancement of pathway FFIS with electronic correlation being roughly absent. This pathway  has not been previously identified. Pathway FFIS with electronic correlation,  identified in our studies with linear fields \cite{Emmanouilidou,Emmanouilidou1,Emmanouilidou2}, still prevails at different ratios of the field strengths. Its main trace is a  ``plateau" in the FDI probability as a function of the ratio of the two field-strengths. 
Moreover, we have developed a simple model to  explain and predict how the FDI probabilities change with  the ratio of the field strengths in CRTC fields. Future studies can explore whether significant increase of frustrated double ionization can be achieved in multi-center molecules, as the current comparison of the FDI probability between D$_{3}^{+}$ and H$_{2}$ implies.
 
A. E. acknowledges  the EPSRC grant no. N031326 and the use of the computational resources of Legion at UCL. Moreover, A. E.  is grateful to Paul Corkum for useful discussions.

%%%
%%%
\bibliography{deuterium_bibliography}{}

\end{document}